# FUZZY BASED STATOR FLUX OPTIMIZER DESIGN FOR DIRECT TORQUE CONTROL

Fatih Korkmaz[1], M. Faruk Çakır[1], Yılmaz Korkmaz[2], İsmail Topaloğlu[1]

[1]Technical and Business Collage, Çankırı Karatekin University, 18200, Çankırı, Turkey
fkorkmaz@karatekin.edu.tr, mcakir@karatekin.edu.tr
,itopaloglu@karatekin.edu.tr
[2]Faculty of Technology, Gazi University, Besevler, 06200, Ankara, Turkey
ykorkmaz@gazi.edu.tr

***ABSTRACT***

*Direct Torque Control (DTC) is well known as an effective control technique for high performance drives in a wide variety of industrial applications and conventional DTC technique uses two constant reference value: torque and stator flux. In this paper, a new fuzzy based stator flux optimizer has been proposed for DTC controlled induction motor drivers and simulation studies have been carried out with Matlab/Simulink to compare the proposed system behaviours at vary load conditions. The most important feature of the proposed fuzzy logic based stator flux optimizer that it self-regulates the stator flux reference value using the motor load situation without need of any motor parameters. Simulation results show that the performance of the proposed DTC technique has been improved and especially at low-load conditions torque ripples are greatly reduced with respect to the conventional DTC.*

## 1. INTRODUCTION

Vector controls are well-known control techniques in high performance variable speed - load applications for induction motor drivers and can be grouped as field oriented control (FOC) and direct torque control (DTC). The basic idea of vector control is the control of motor flux and torque separately as dc motors. For this aim, motor currents converted two phase vector components using park or Clarke transformations. One of these is the component control flux vector, and the other one is the control torque vector, separately. The main difference between the two methods is that the FOC controls by a rotor or stator field orientation while the DTC controls by stator field observation[1].

FOC was first introduced by Blaschke[2] in the 1970's. It was unrivalled in industrial induction motor drivers until DTC was introduced by Takahashi[3] in the middle of the 1980's. It was a good alternative to FOC due to some well known advantages, such as simple control structure, no need much motor parameters so independecy of parameter changes, fast dynamic response. Beside these advantages, DTC scheme still had some disadvanteges like high torque and current ripples, variable switching frequency behavior and implementaiton difficulties owing to necessity of low sampling time.

Over the years, many studies have been done to overcome these disadvantages of the DTC and still continues. In this studies, researchers proposed different ways. Some studies suggest using different switching techniques and inverter topologies [4-5], in other group of researches, different observer models have been suggested [6-7]. On the other hand, intelligent control methods like fuzzy logic (FL) have been explored by several researchers for its potential to improve the speed regulation of the drive system. [8-9]

On the other hand, intelligent control methods like fuzzy logic have been explored by several researchers for its potential to incorporate human intuition in the design process. The FL has gained great attention in the every area of electromechanical devices control due to no need mathematical models of systems unlike conventional controllers[10]. A fuzzy logic controller is used to select voltage vectors in conventional DTC in some applications [11-13]. In parameter estimation applications, a fuzzy logic stator resistance estimator is used and it can estimate changes in stator resistance due to temperature change during operation[14]. For duty ratio control method, a fuzzy logic controller is used to determine the duration of output voltage vector at each sampling period [15]. These fuzzy logic controllers can provide good dynamic performance and robustness.
In recent publications, we see that some flux optimization method is proposed for the DTC scheme for induction motor drives and the effects of the optimization algorithm is investigated. In these publications, three flux control methods are used for optimization and essentially we can classified according to control structure as following: flux control as a function of torque [16], flux control based on loss model [17] and flux control by a minimum loss search controller [18].

This paper deals with a new stator flux controller on the DTC scheme. It has been developed to determine the best flux reference value for motor using the FL algorithm. The proposed controller self-regulates the stator flux reference without need of any motor parameters. Simulation studies have shown that this method reduces the torque ripple of the DTC scheme.

## 2. DIRECT TORQUE CONTROL

The basic idea of the DTC technique, which its block diagram is shown in Fig. 1, is to choose the best vector of the voltage, which makes the flux rotate and produce the desired torque. In the DTC motor drive, the stator flux linkage and the electromagnetic torque can be directly controlled by the selection of optimum inverter switching states. The flux and the torque errors are kept within acceptable limits by hysteresis controllers. The DTC allows for very fast torque responses, and flexible control of the induction motor[19]

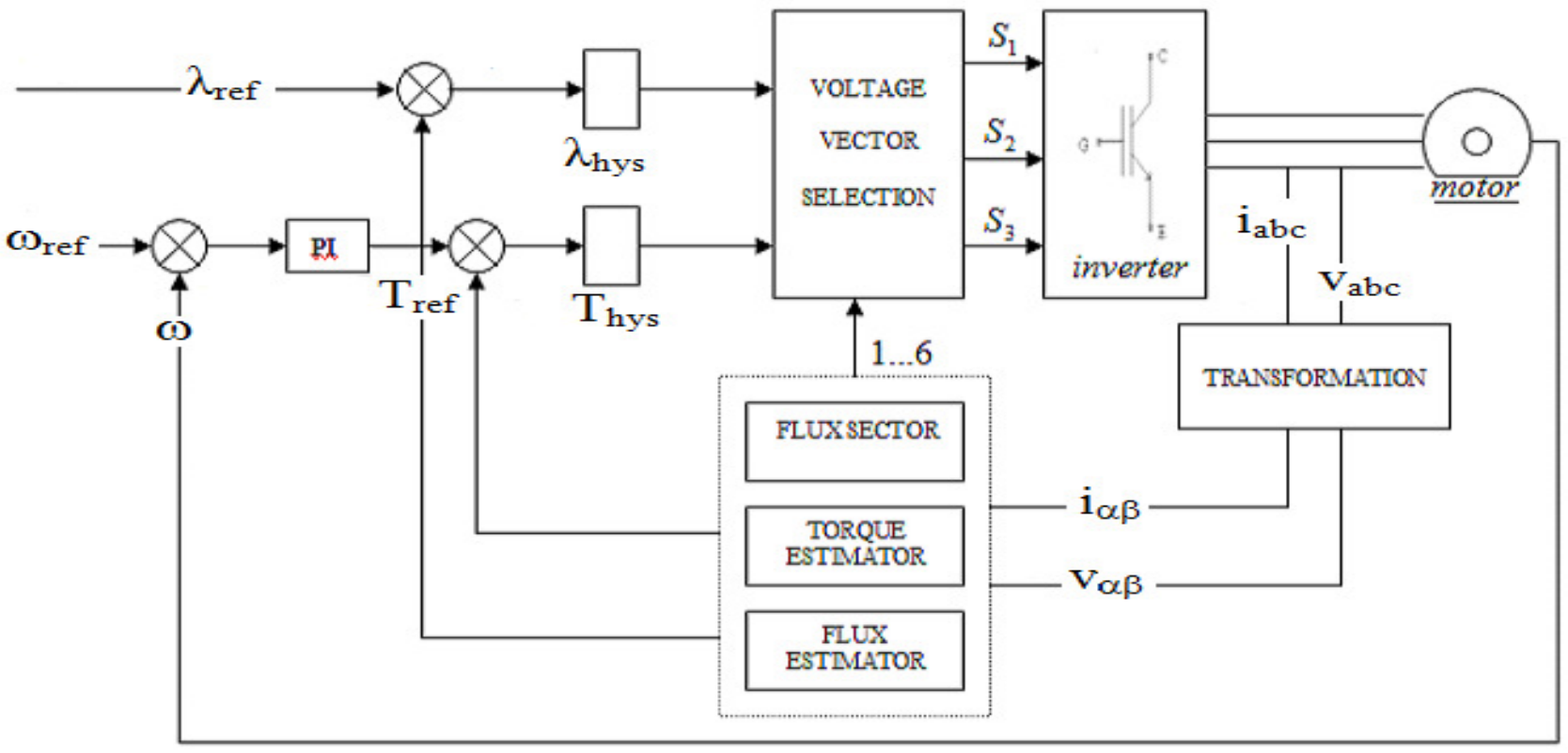

Figure 1. Block diagram of conventional DTC

Instantaneous values of the flux and the torque are calculated by using the transformation of the measured currents and the voltages of the motor. In these calculations, all measured electrical values of motor must be converted to stationary α-β reference frame on DTC scheme and conversation matrix as given in (1-3).

$$i_{\alpha\beta 0} = [T]\ i_{abc} \tag{1}$$

$$V_{\alpha\beta 0} = [T]\ V_{abc} \tag{2}$$

$i_{abc}$ , $v_{abc}$ measured and $i_{\alpha\beta 0}$ , $v_{\alpha\beta 0}$ calculated phase currents and voltages respectively. $T$ is transformation matrix as given in (3).

$$T = \frac{2}{3}\begin{bmatrix} 1 & -\frac{1}{2} & -\frac{1}{2} \\ 0 & -\frac{\sqrt{3}}{2} & \frac{\sqrt{3}}{2} \\ \frac{1}{2} & \frac{1}{2} & \frac{1}{2} \end{bmatrix} \tag{3}$$

Stator flux vector can be calculated using the measured current and voltage vectors as given in (4-6).

$$\lambda_\alpha = \int (V_\alpha - R_s i_\alpha) dt \tag{4}$$

$$\lambda_\beta = \int (V_\beta - R_s i_\beta) dt \tag{5}$$

$$\lambda = \sqrt{\lambda_\alpha^2 + \lambda_\beta^2} \tag{6}$$

Where $\lambda$ is stator flux space vector, $v_{ds}$ and $v_{qs}$ stator voltage, $i_{ds}$ and $i_{qs}$ line currents in α-β reference frame and $R_s$ stator resistance. The electromagnetic torque of an induction machine is usually estimated as given in (7).

$$T_e = \frac{3}{2} p(\lambda_\alpha i_\beta - \lambda_\beta i_\alpha) \tag{7}$$

Where $p$ is the number of pole pairs. An important control parameter on DTC is stator flux vector sector. Stator flux rotate trajectory divided six sector and calculation of stator flux vector sector as given in (8).

$$\theta_\lambda = \tan^{-1}(\frac{\lambda_\beta}{\lambda_\alpha}) \tag{8}$$

Two different hysteresis comparator generates other control parameters on DTC scheme. Flux hysteresis comparator is two level type while torque comparator is tree level type. These comparators use flux and torque instantaneous error values as input and generates control signals as output. Switching selector unit generates inverter switching states with use of hysteresis comparator outputs and stator flux vector sector.

## 3. FUZZY FLUX OPTIMIZATION BASED DTC SYSTEM

As can be seen in Fig.1., conventional DTC scheme not only uses speed/torque reference value, but also a stator flux reference value as control parameters. Usually, motors are designed to work their maximum efficiency in their nominal operating point. However, for many industrial control applications motor loading situations can vary from time to time. So, it has been investigated in this paper what if flux reference value is adjusted regarding to the motor load. The value of motor flux should be readjusted when the load is less than the rated value.

Adaptation of flux to load variations can be done in three ways: flux control as a function of torque, flux control based on loss model and flux control by a minimum loss search controller. In this paper, the first way have been preferred. It means that flux controlled as a function of torque but without need of any motor parameters by using fuzzy algorithm. Simulink block diagram of the proposed DTC scheme is given in Fig. 2.

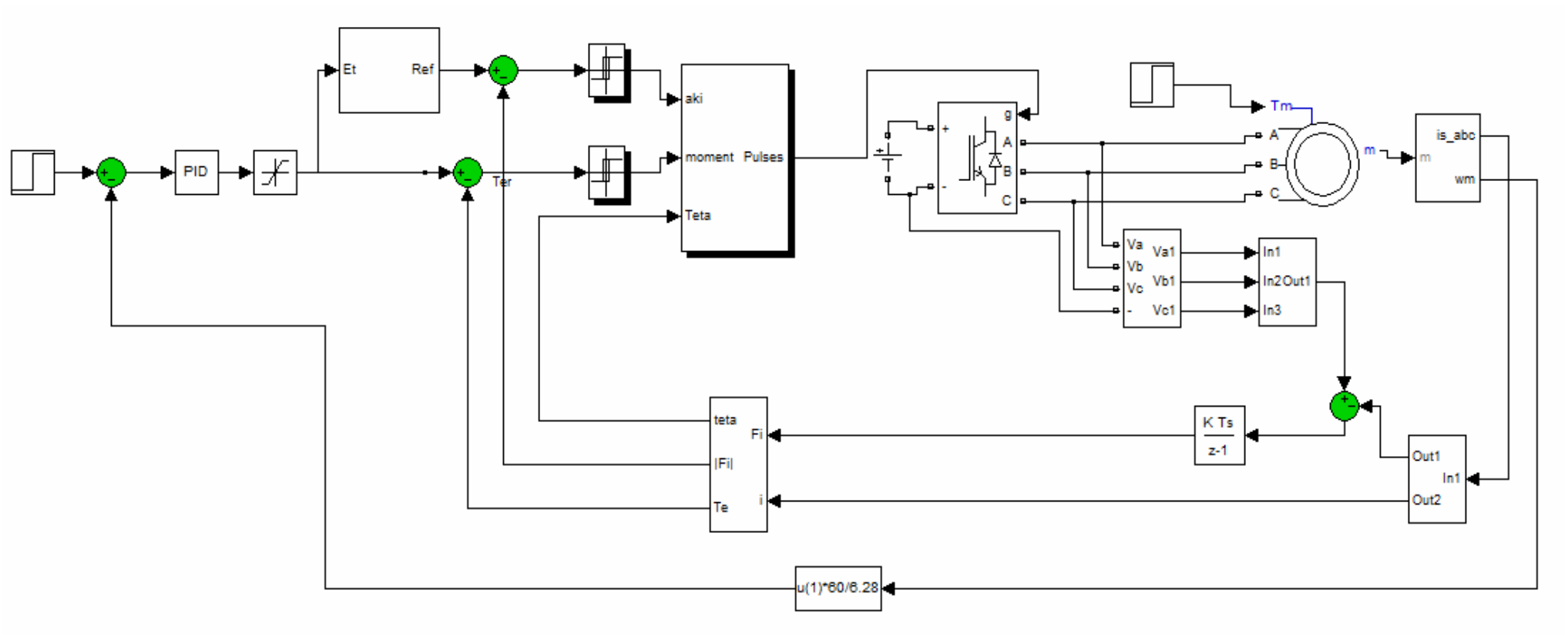

Figure 2. Simulink block diagram of proposed DTC

FL controller, which used in proposed DTC scheme, utilizes the torque error and initial value of stator flux reference as control variable and generates amount of change on stator flux reference for next step as output.

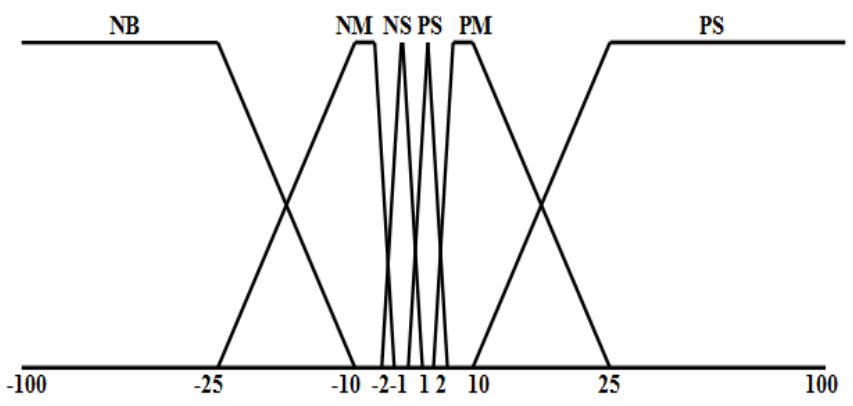

(a) Torque error

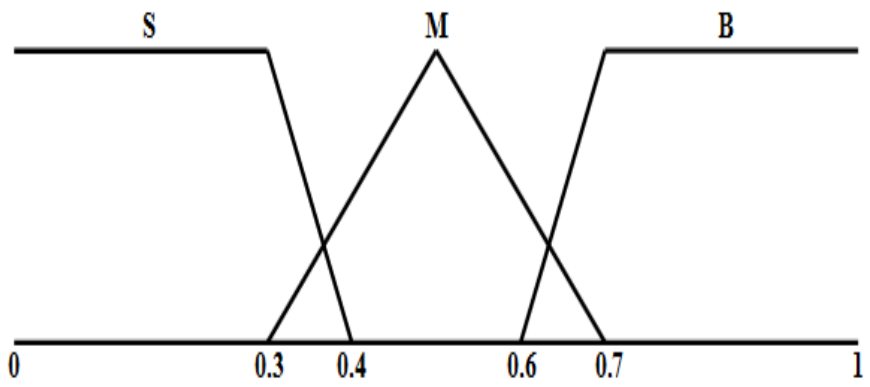

(b) Initial value of stator flux reference

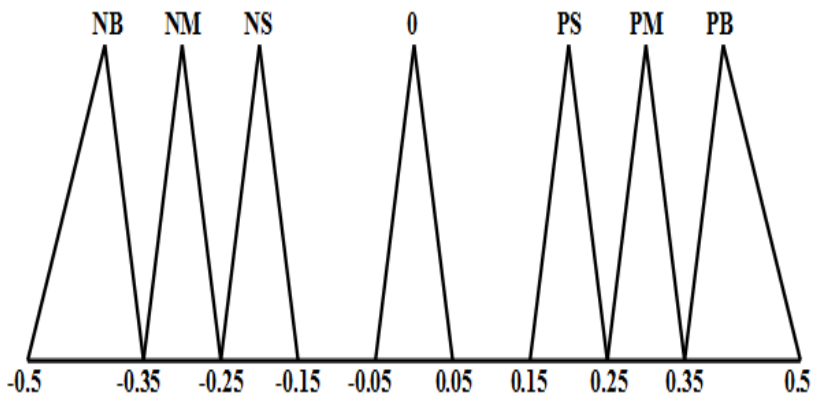

(c) Change on stator flux reference
Figure. 3. Membership functions

Membership functions of purposed fuzzy control scheme are given in Fig 3. Fuzzy control rules of purposed fuzzy control scheme are designed to minimize torque ripples and rules can be obtained based on prior experience of investigators about DTC scheme. The FL rules shown in Table 1.

Table 1. Rule Table

| | | Torque error | | | | | |
|---|---|---|---|---|---|---|---|
| | | NB | NM | NS | PS | PM | PB |
| $\lambda_{ref}^{-1}$ | S | 0 | PS | PB | 0 | PS | PB |
| | M | NS | 0 | PM | NS | 0 | PM |
| | B | NB | NM | 0 | NB | NM | 0 |

## 4. SIMULATIONS

Numerical simulations have been carried out to investigate the effects the proposed fuzzy stator flux controller based DTC scheme. Its developed using Matlab/Simulink®. The parameter of the induction motor and simulation used in research as follows:

Table 2. Parameters of Motor and Simulations

| | |
|---|---|
| Rated Power (kW) | 7.5 |
| Rated Voltage (V) | 400 |
| Frequency (Hz) | 50 |
| Rated speed (rpm) | 1440 |
| Stator Resistance (Ω) | 0.7334 |
| Pole pairs (*p*) | 2 |
| DC bus voltage (V) | 400 |
| Reference speed (rpm) | 1500 |
| Cycle period (μs) | 50 |

In the first step of simulation studies, motor has worked with 0 N.m. reference torque value to compare unloaded behaviors of the motor on conventional and proposed DTC. The torque response curves of conventional DTC and proposed fuzzy stator flux optimizer based DTC are shown Fig 4.

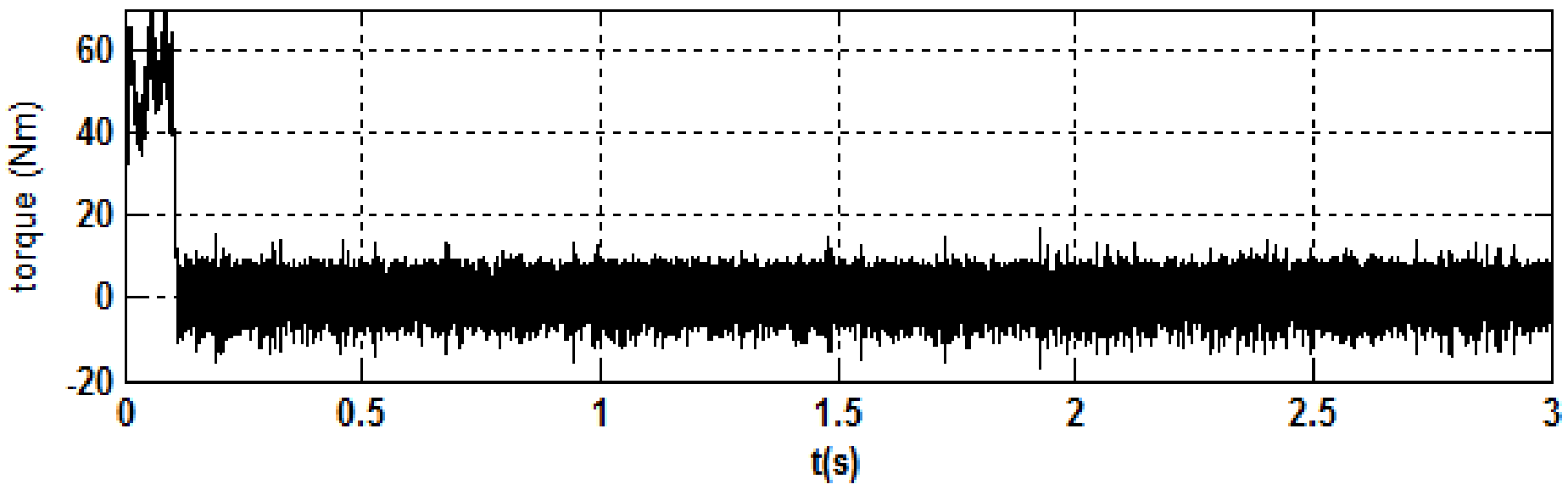

Conventional DTC

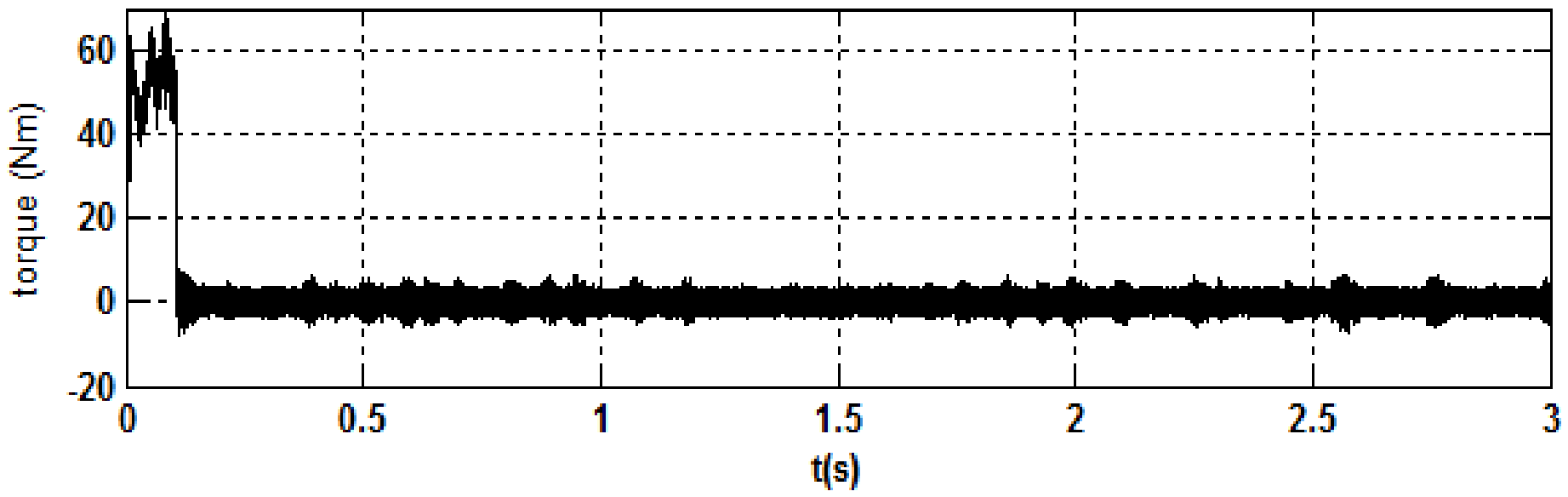

Proposed DTC

Figure 4. Torque response of conventional DTC and proposed DTC

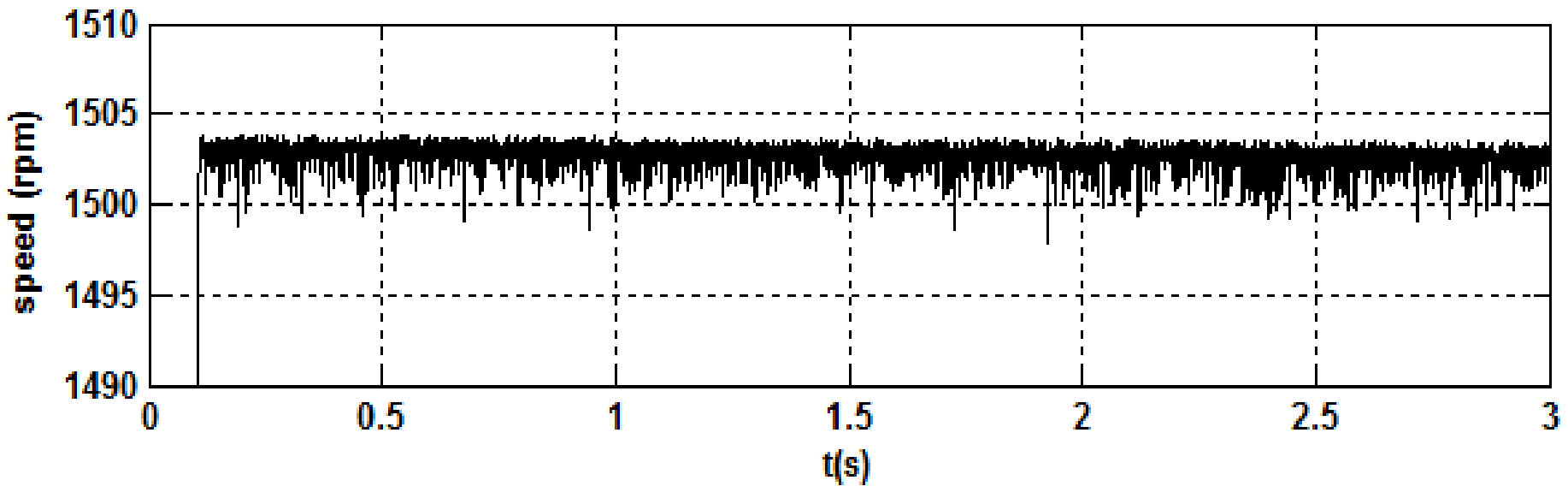

Conventional DTC

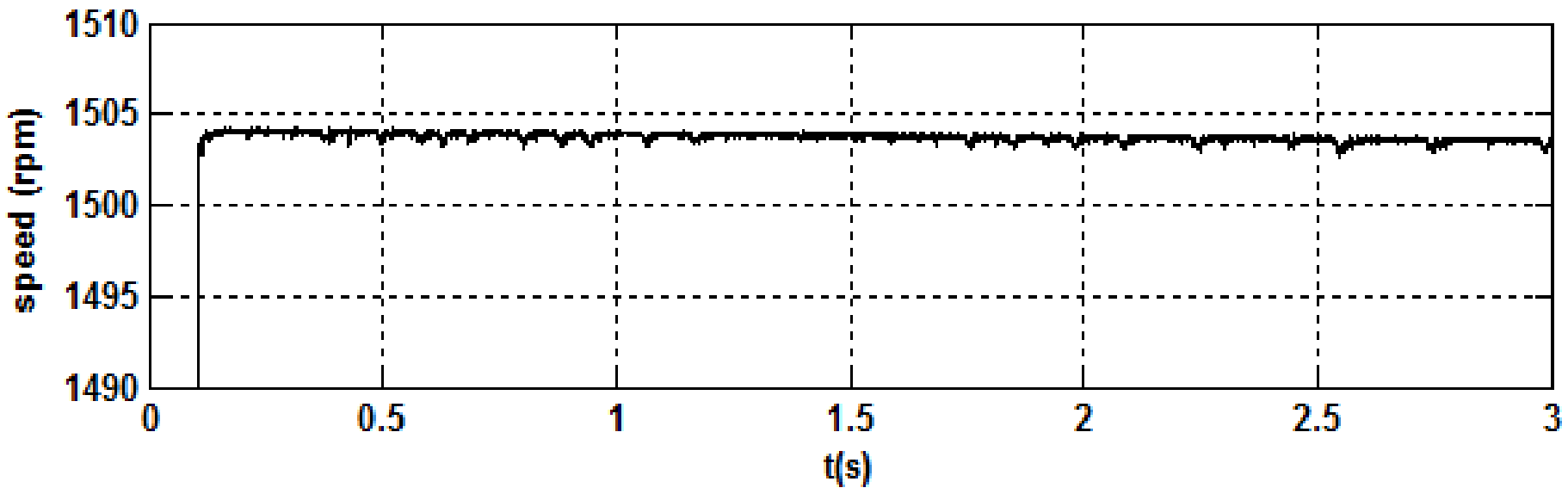

Proposed DTC

Figure 5. Speed response of conventional DTC and proposed DTC

It can be seen that the proposed stator flux optimization system finds the optimal flux value rapidly and has better performance. Obviously, proposed system with optimized command stator flux has much smaller ripple in torque with respect to the conventional DTC at all working conditions. When the speed responses of motor are compared, its clear that the motor speed ripples has reduced remarkably with proposed DTC controlled motor. The speed response curves of motor with unloaded conditions are given Fig. 5.

In the next step of simulation studies, motor has worked with 15 N.m. reference torque value to compare loaded behaviors of motor on conventional and proposed DTC. The torque and speed response curves of conventional DTC and proposed fuzzy stator flux optimizer based DTC are shown Fig 6. and Fig. 7.

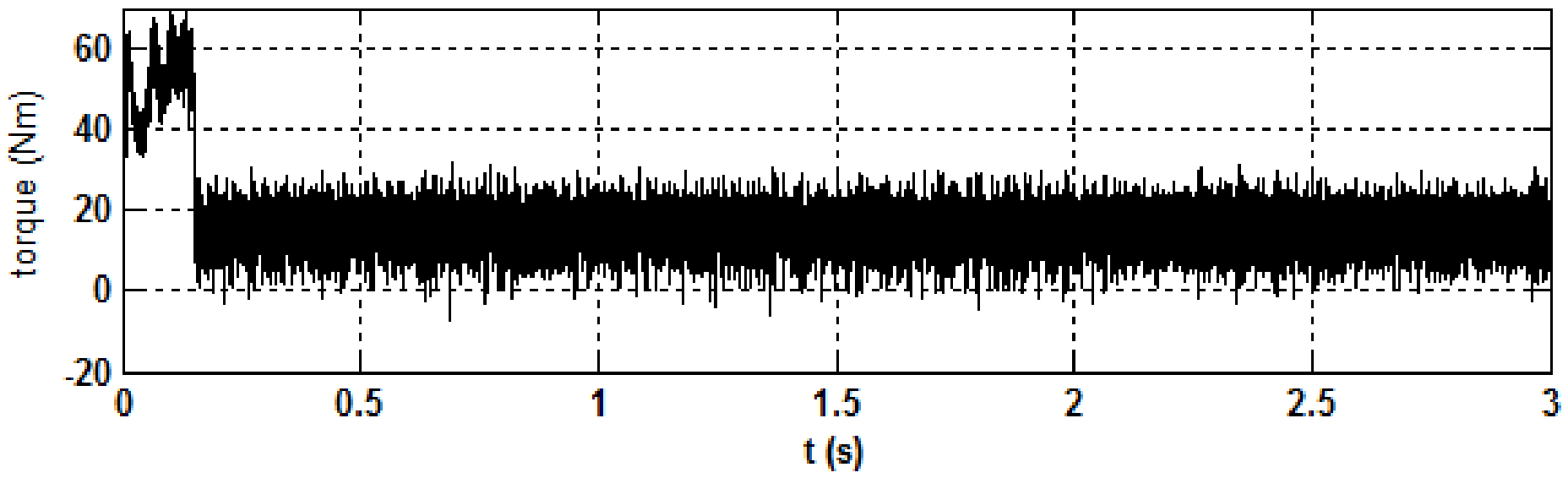

Conventional DTC

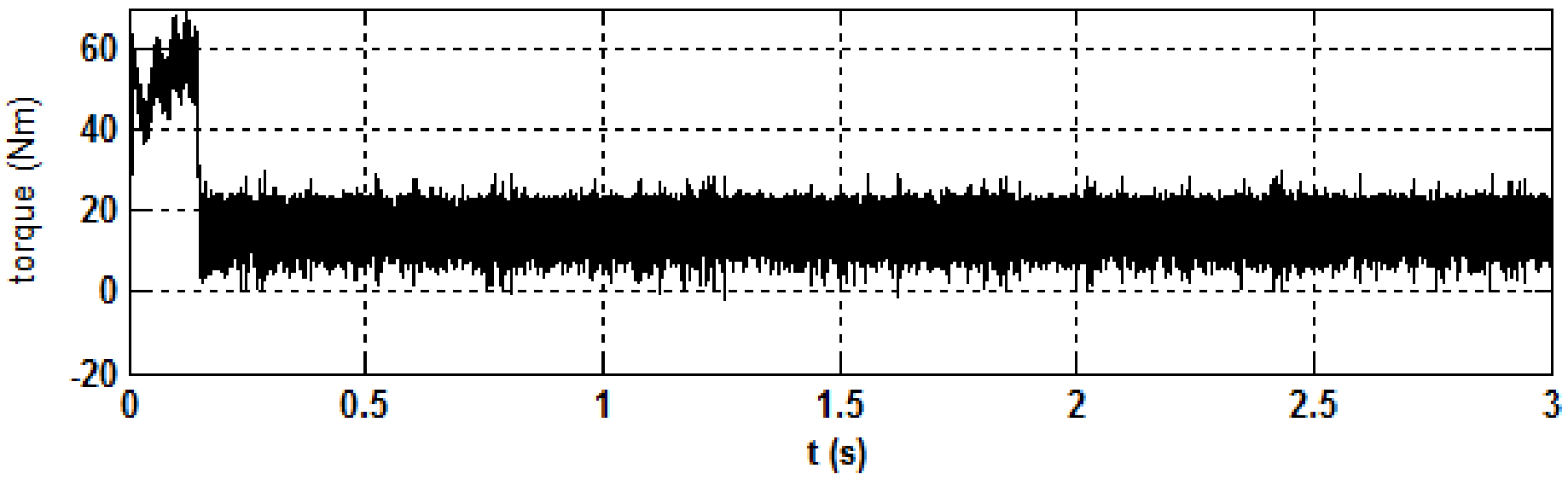

Proposed DTC

Figure 6. Torque response of conventional DTC and proposed DTC

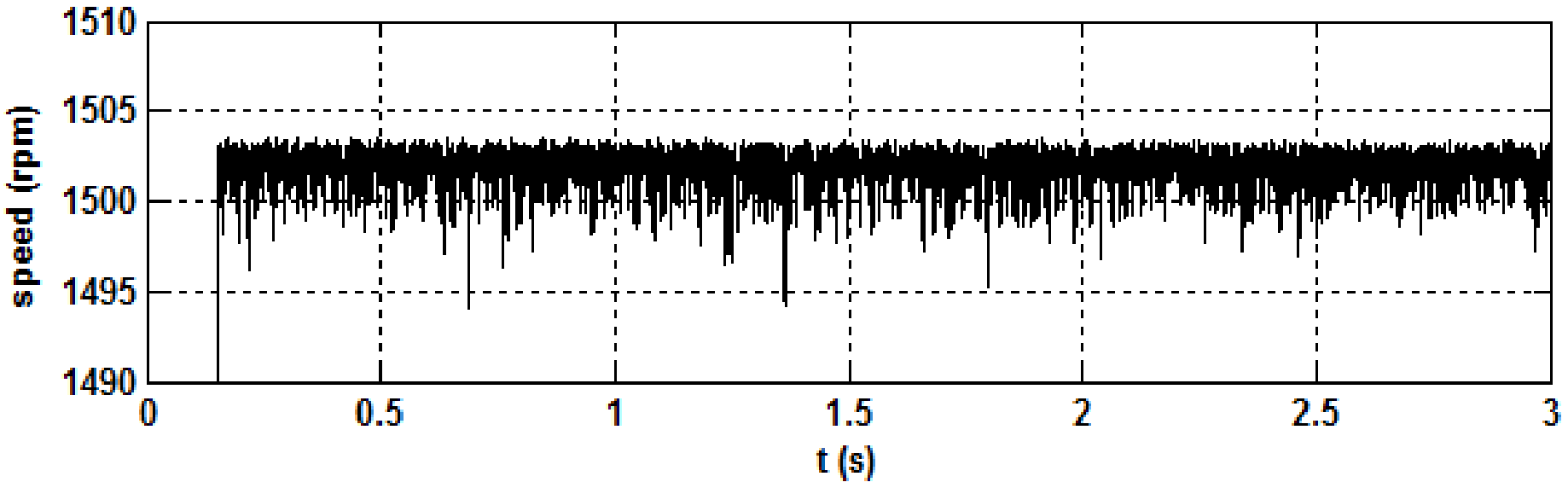

Conventional DTC

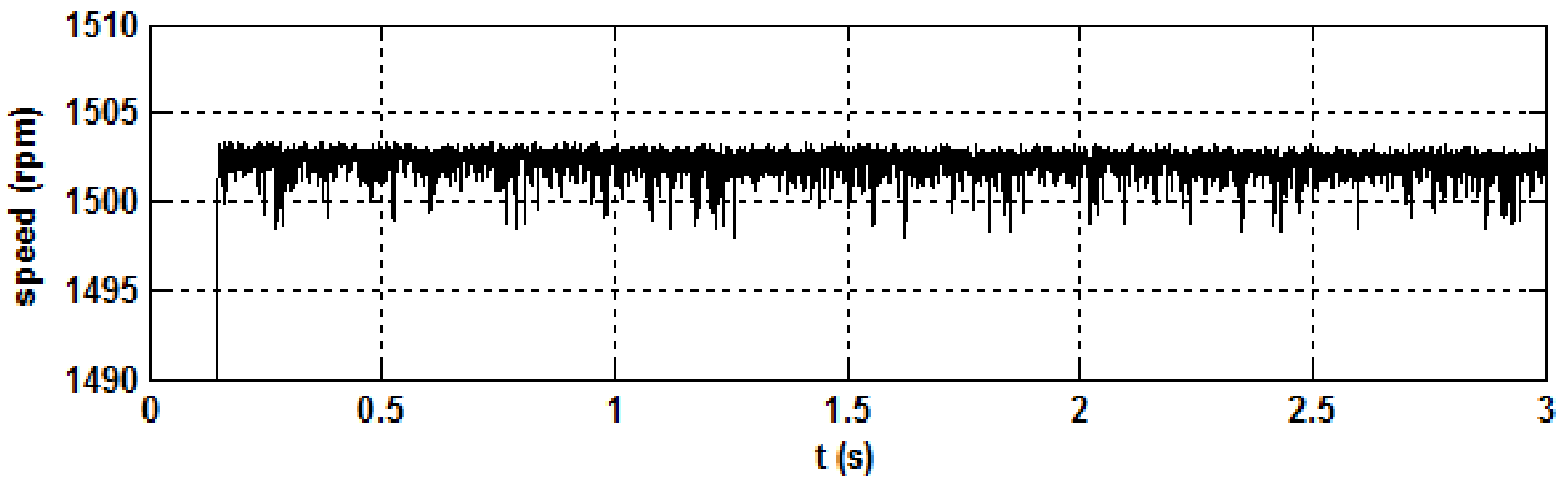

Proposed DTC

Figure 7. Speed response of conventional DTC and proposed DTC

The torque and speed response curves show that proposed DTC controlled motor still has lesser torque and speed ripples regarding to conventional DTC controlled one. Transient working behaviors of both control technique are completely same due to optimizer unit works based instantaneous torque error and the torque error value is at largest value in transient working conditions. So, at transient working, the fuzzy optimizer unit creates rate flux value which used in conventional DTC.

## 6. CONCLUTION

A new fuzzy logic based stator flux control and optimization strategy for DTC controlled induction motors has been presented and investigated in this paper. Fuzzy logic based stator flux optimizer has been designed to determine the reference value. Fuzzy stator flux optimizer uses just torque error and change on torque error without need of any other motor parameter to determine the reference value on the DTC scheme. The proposed DTC system structure also has resistant to changes in motor parameters as conventional DTC due to use of any additionally motor parameters. So, the proposed DTC system can easily applicable every size of motors. The simulation results validate that fuzzy logic based control strategy for stator flux optimization can be successfully cooperated with conventional DTC scheme and achieves a reduction of torque ripple.

## 7. REFERENCES

[1] Liu-Jun; Wang Wan-li; Wang Yang; , "Research on FOC and DTC switching control of asynchronous motor based on parameter estimation," Automation and Logistics, 2008. ICAL 2008. IEEE International Conference on , vol., no., pp.1754-1758, 1-3 Sept. 2008

[2] Blaschke, F., “The Principle of Field Orientation Applied to The New Transvector Closed-Loop Control System for Rotating Field Machines”, Siemens-Rev., Vol. 39, 217–220, 1972.

[3] Takahashi, I., Noguchi. T., “A new quick-response and high efficiency control strategy of an induction motor,” IEEE Transactions on Industrial Applications, 1986, vol.I A-22 ,No.5. , pp. 820–827.

[4] D. Casadei, G. Serra and A. Tani, “The use of matrix converters in direct torque control of induction machines,” IEEE Trans. on Industrial Electronics, 2001, vol.48 , no.6 , pp. 1057–1064.

[5] D. Casadei, G. Serra and A. Tani, “Implentation of a direct torque control algorithm for induction motors based on discrete space vector modulation,” IEEE Trans. on Power Electronics, 2000, vol.15 , no. 4 , pp. 769–777.

[6] Z. Tan, Y. Li and Y. Zeng, “A three-level speed sensor-less DTC drive of induction motor based on a full-order flux observer,” Power System Technology, Proceedings. PowerCon International Conference, 2002, vol. 2, pp. 1054- 1058.

[7] G. Ya and L. Weiguo, "A new method research of fuzzy DTC based on full-order state observer for stator flux linkage," Computer Science and Automation Engineering (CSAE), 2011 IEEE International Conference, 2011, vol. 2 , pp.104-108.
[8] S. Benaicha, F. Zidani, R.-N. Said, M.-S.-N. Said, "Direct torque with fuzzy logic torque ripple reduction based stator flux vector control," Computer and Electrical Engineering, (ICCEE '09), 2009, vol.2 , pp. 128–133.
[9] N. Sadati, S. Kaboli, H. Adeli, E. Hajipour and M. Ferdowsi, "Online optimal neuro-fuzzy flux controller for dtc based induction motor drives," Applied Power Electronics Conference and Exposition (APEC 2009), 2009, pp.210–215.
[10] Li, H., "Fuzzy DTC for induction motor with optimized command stator flux," Intelligent Control and Automation (WCICA), 2010, pp. 4958–4961.
11] Bacha, F., Dhifaoui, R., Buyse, H., Real-time implementation of direct torque control of an induction machine by fuzzy logic controller, International Conference on Electrical Machines and Systems (ICEMS),2001. – Vol. 2, P. 1244–1249.
[12] Mir, S., Elbuluk, M.E., Zinger, D.S., PI and fuzzy estimators for tuning the stator resistance in direct torque control of induction machines, IEEE Trans. Power Electron. 1998. – Vol.13 – P. 279–287.
[13] Arias, A., Romeral, J.L., Aldabas, E., Jayne, M.G., Fuzzy logic direct torque control, IEEE International Symposium on Industrial Electronics (ISIE), 2000. – Vol.1 – P. 253–258.
[14] Holtz, J., Quan, J., Sensorless vector control of induction motors at very low speed using a nonlinear inverter model and parameter identification, IEEE Trans. Ind. Appl., 2002. Vol. 38 – No. 4 – P. 1087–1095.
[15] Casadei, D., Serra, G., Tani, A., Zarri, L., Profumo, F., Performance analysis of a speed-sensorless induction motor drive based on a constant switching-frequency DTC scheme, IEEE Trans. Ind. Appl., 2003. – Vol. 39 – No. 2 – P. 476–484.
[16] Kaboli, S., Zolghadri, M.R., Haghbin, S.,Emadi, A., "Torque ripple minimization in DTC of induction motor based on optimized flux value determination," IEEE Ind. Electron. Conf., 2003, pp. 431–435.
[17] Kioskeridis, I., Margaris, N., "Loss minimization in induction motor adjustable speed drives," IEEE Trans. Ind. Electron., 1996, vol. 43, no. 1, pp. 226–231.
[18] Kioskeridis, I., Margaris, N., "Loss minimization in scalar-controlled induction motor drives with search controllers," IEEE Trans. Power Electron., 1996, vol. 11, no. 2, pp. 213–220.
[19] P. Vas, Sensorless vector and direct torque control. Oxford UniversityPress,2003

# AUTHORS

Fatih Korkmaz was born in Kırıkkale, Turkey in 1977. He received the B.T., M.S., and Doctorate degrees in in electrical education, from University of Gazi, Turkey, respectively in 2000, 2004 and 2011. His current research field includes Electric Machines Drives and Control Systems.

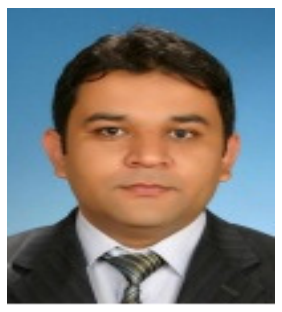

M. Faruk Çakır was born in Turkey in 1972. He received B.T. degree from depertmant of electric-electronic engineering, Selçuk University, Turkey, in 1994 and M.S. degree from Gebze High Technology Institute , Turkey, in 1999. Now he is PhD student in University of Gazi, Turkey. His research deals with Electric Machine Design and Nano Composite Materials.

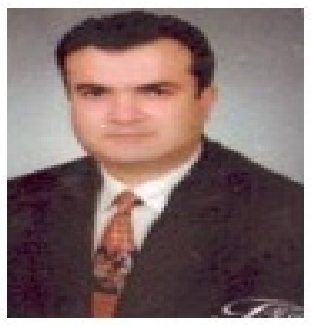

Yılmaz Kormaz was born in Çorum, Turkey in 1956. He received the B.T., M.S., and Doctorate degrees in electrical education from University of Gazi, Turkey, respectively in 1979, 1994 and 2005. His current research field includes Electric Machines Design and Control.

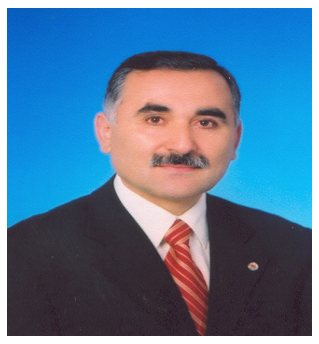

İsmail TOPALOĞLU was born in Adana, Turkey, in 1983. He received the B.Sc and M.Sc. degrees in electrical education from University of Gazi in 2007 and 2009, respectively. His current research interests include Computer aided design and analysis of conventional and novel electrical and magnetic circuits of electrical machines, sensors and transducers, mechatronic systems.

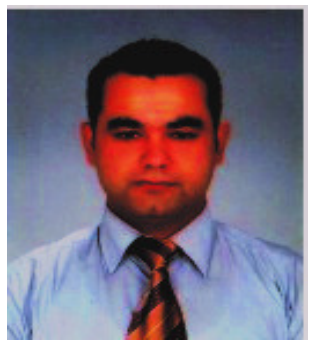